\documentclass[%
 reprint,
superscriptaddress,
 amsmath,amssymb,
 aps,
 prl
]{revtex4-1}

\usepackage{graphicx}
\usepackage{dcolumn}
\usepackage{bm}
\usepackage{hyperref}

\begin{document}

\preprint{APS/123-QED}

\title{Application of Jordan Decomposition to Non-Hermitian Lattice Models
with Spectrally-Isolated Lower Dimensional States}

\author{Yang Yu}
 \affiliation{School of Applied and Engineering Physics, Cornell University, Ithaca, NY 14853, USA}
\author{Minwoo Jung}
 \affiliation{Department of Physics, Cornell University, Ithaca, NY 14853, USA}
\author{Gennady Shvets}%
 \email{gshvets@cornell.edu}
 \affiliation{School of Applied and Engineering Physics, Cornell University, Ithaca, NY 14853, USA}
\date{\today}

\begin{abstract}
When analyzing non-Hermitian lattice systems, the standard eigenmode decomposition utilized for the analysis of Hermitian systems must be replaced by Jordan decomposition. This approach enables us to identify the correct number of the left and right eigenstates of a large finite-sized lattice system, and to form a complete basis for calculating the resonant excitation of the system. Specifically, we derive the procedure for applying Jordan decomposition to a system with spectrally-isolated states. We use a non-Hermitian quadrupole insulator with zero-energy corner states as an example of a large system whose dimensionality can be drastically reduced to derive a low-dimensional ``defective" Hamiltonian describing such localized states. Counter-intuitive and non-local properties of the resonant response of the system near zero energy are explained using the Jordan decomposition approach. Depending on the excitation properties of the corner states, we classify our non-Hermitian quadrupolar insulator into three categories: trivial, near-Hermitian, and non-local.
\end{abstract}

\maketitle


Non-Hermitian physics has attracted considerable interest in recent years because of its relevance to non-equilibrium (e.g., undergoing photo-ionization) systems~\cite{baker_pra84,lopata_jctc13}. Some of its notable phenomena include ``exceptional points" (EPs)~\cite{heiss2004exceptional,berry2004physics,moiseyev2011non} and real-valued spectra despite non-Hermiticity. At the EP, both the complex-valued eigenvalues of two bands as well as their corresponding eigenvectors coalesce~\cite{liang_feng_nphot17,el2018non}. In other words, the matrix corresponding to the Hamiltonian at the EP becomes {\it defective} \cite{golub2013matrix,lee2016anomalous}. The completely real spectrum of some non-Hermitian systems can be related to parity-time (PT) symmetry \cite{bender2007making,PhysRevLett.104.054102,PhysRevA.88.062111} or pseudo-Hermiticity \cite{mostafazadeh2002pseudo}, though in general it is hard to assert a real spectrum without directly calculating the eigenvalues.

In this paper we will focus on non-Hermitian lattice models, where the Hamiltonian of the system can be represented by a finite dimensional non-Hermitian matrix. There has been a considerable amount of literature on these systems recently \cite{liu2019second,lee2016anomalous,PhysRevLett.121.026808,yao2018edge,PhysRevB.98.165148,longhi2015robust,PhysRevX.8.041031,PhysRevResearch.2.013058,PhysRevB.99.081103}. However, most literature uses eigendecomposition, which is in fact only appropriate for Hermitian matrices \cite{golub2013matrix}. This leads to at least two possible issues: (1) not identifying all approximate eigenstates of the system and (2) the eigenstates do not form a complete basis. To see why (1) can happen, consider the following two-level Hamiltonian: 
\begin{equation}
    H=
    \begin{pmatrix}
    0 & q_1^N\\
    q_2^N & 0
    \end{pmatrix},
\end{equation}
in the basis $\{|1\rangle,|2\rangle\}$, where $|q_2|<|q_1|<1$ and $N\gg1$ is the size of the system. The usual eigendecomposition would give the eigenstates of the system as $|1\rangle\pm(q_2/q_1)^{N/2}|2\rangle\approx|1\rangle$ when $N\to\infty$. However, in reality, the exponentially small off-diagonal terms $q_1^N$ and $q_2^N$ should not be compared against each other. Rather, they should be compared against some characteristic energy scale $E_c$ of the system. If $|q_1|^N\ll E_c$, then states $|1\rangle$ and $|2\rangle$ are both good approximate eigenstates of the system. For example, if we observe the system on a time scale much shorter than $1/|q_1|^N$, then if the system starts in either state it will remain in that state during our observation. Likewise, if the system has some finite loss $\Gamma$, which is inevitable for any real-world passive systems, then $E_c\sim\Gamma$. A third scenario is when the states $|1\rangle$ and $|2\rangle$ have a small but finite energy split $\epsilon$, which is also inevitable for realistic systems, then $E_c\sim\epsilon$. This example demonstrates that the eigendecomposition of non-Hermitian matrices can be numerically unstable and return pathological results. The second issue comes into play in the excitation of modes and can lead to the ``missing dimension" \cite{chen2020revealing}.

There are many decomposition that are applicable to non-Hermitian matrices but the one most relevant to the physics community and will serve as the replacement for the eigendecomposition is the Jordan decomposition. This is because we are interested in solving the equation of motion (EOM) of the system: $id\psi/dt=H\psi+\xi$, either without a source ($\xi=0$, homogeneous) or with a source ($\xi\neq0$, inhomogeneous), which involves calculating the resolvent or exponential of the Hamiltonian, and the resolvent or exponential of a Jordan block is easy to calculate (this is also the primary reason why we use the eigendecomposition for Hermitian matrices). For example, suppose we have a periodic source with frequency $E$, and the spectrum of the system is purely real (either because of PT symmetry or pseudo-Hermiticity), then adding a small overall loss $\Gamma$ to the system ensures all transients eventually  decay. Therefore, only the driven equation $(E-H+i\Gamma)\psi=\xi$ needs to be solved. If we obtain the Jordan decomposition of the Hamiltonian $H=PJP^{-1}$, we can instead solve $(E-J+i\Gamma)\psi'=\xi'$ where $\xi'=P^{-1}\xi,\psi=P\psi'$. We see we need to obtain both the relevant columns of $P$ matrix and rows of $P^{-1}$ corresponding to eigenvalues near $E$. We call the former right Jordan basis and the latter left Jordan basis, a generalization of right and left eigenvectors.

While it is known that Jordan decomposition is also numerically unstable, the Schur decomposition is \cite{golub2013matrix}. After obtaining the Schur decomposition, one can group close eigenvalues together and obtain invariant subspaces. In practice this can be difficult for bulk continuum of the spectrum, but if there are lower dimensional modes that are isolated in the spectrum, this approach is straightforward. What we will get is a block diagonal form
\begin{equation}
    H=P
    \begin{pmatrix}
    J_0 & 0\\
    0 & J_r\\
    \end{pmatrix}
    P^{-1},
\end{equation}
where $J_0$ is a $n\times n$ upper diagonal matrix with grouped eigenvalues $E_0$ on the main diagonal. The explicit procedures are: (1) calculate the Schur decomposition: $H=UTU^\dagger$, and find isolated eigenvalues we are interested in; (2) order the Schur decomposition so that these eigenvalues appear first on the main diagonal of $T$, then the first block of $T$ is $J_0$, and the first $n$ columns of $U$, $\eta_1^R,\cdots,\eta_n^R$, are the first $n$ columns of $P$; (3) repeat the above steps on $H^T$, get the first columns of $U$ (a different one) as $\eta_1^L,\cdots,\eta_n^L$; (4) let $\Sigma=(\eta_1^L,\cdots,\eta_n^L)^T(\eta_1^R,\cdots,\eta_n^R)$, then $\Sigma^{-1}(\eta_1^L,\cdots,\eta_n^L)^T$ are the first $n$ rows of $P^{-1}$. 

\begin{figure}
\includegraphics[width=\linewidth]{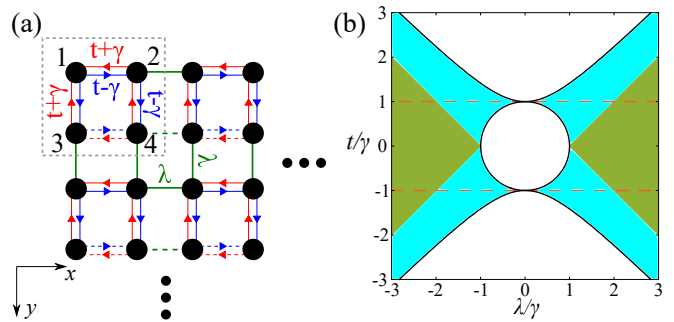}
\caption{\label{fig:nhqi} (a) Tight binding model of a non-Hermitian QI on a square lattice. Grey dashed line: boundary of unit cell with four (sublattice) sites (numbered 1 to 4). Red and blue lines with arrows: asymmetric intra-cell hopping amplitudes $\pm t \pm \gamma$, green lines: symmetric inter-cell hopping amplitudes $\pm \lambda$. Dashed lines: negative hopping terms. All four sublattices have the same on-site potentials (set to $\epsilon_j \equiv 0$).
(b) The phase diagram of a large non-Hermitian QI with open boundary condition. Green region ($|\lambda| > |t| + |\gamma|$): near-Hermitian regime with $4$ zero-energy corner states, each localized at a separate corner. Cyan region ($\sqrt{|t^2-\gamma^2|} <|\lambda| < |t|+|\gamma|$): intermediate regime with $2$ zero-energy corner states at the top-left corner. White region ($|\lambda| < \sqrt{|t^2-\gamma^2|}$): no corner states. Bandgap vanishes along solid black lines. The spectrum is complex-valued between the two dashed orange lines, real-valued elsewhere.}
\end{figure}

Next we use a non-Hermitian quadrupole insulator (QI) model from Ref.~\cite{liu2019second} as a nice example to demonstrate how Jordan decomposition comes in handy when analyzing the behavior of the system. This model is schematically shown in Fig.~\ref{fig:nhqi}(a), where the intra/inter-cell hopping amplitudes $t\pm\gamma$ and $\lambda$ are all taken to be real. It is a natural non-Hermitian generalization of the QI model described in Ref.~\cite{benalcazar2017quantized}, with the intracell hopping strength becoming asymmetric, characterized by a finite $\gamma$, while maintaining the sublattice symmetry $\Sigma H\Sigma^{-1}=-H$. Here the symmetry operator $\Sigma=P_1-P_2-P_3+P_4$, where $P_j=\sum_{x,y}|x,y,j\rangle\langle x,y,j|$ are the sublattice projection operators, and $|x,y,j\rangle$ are the tight-binding states, where $x$ and $y$ are integer-valued coordinates of the unit cells as defined in Fig.~\ref{fig:nhqi}(a), and $j=1,\dots,4$ denote four sub-lattice sites of each unit cell. This model can also be viewed as a two-dimensional (2D) generalization of the non-Hermitian Su-Schrieffer-Heeger (SSH) model \cite{lieu2018topological,yin2018geometrical,yao2018edge}.

As was pointed in the context of the non-Hermitian SSH system~\cite{yao2018edge}, the open-boundary spectrum can significantly differ from that of the periodic-boundary system described by the Bloch Hamiltonian $H(\vec{k})$. That is because the usual Bloch phase-shift factor $e^{ik}$ for bulk eigenstates (i.e., eigenstates in the continuum spectrum) of an open-boundary system needs to be modified to $\beta\equiv \beta_0e^{ik}$, where $\beta_0$ can be non-unity (i.e., the wavevector acquires an imaginary part: $k\to k-i\ln \beta_0$). This extra \textit{bulk localization factor} $\beta_0$ must be taken into account when calculating the spectrum of the open-boundary system. The same argument applies to our 2D non-Hermitian QI system, where $\vec{k}\equiv(k_x,k_y)\to(k_x-i\ln \beta_0,k_y-i\ln \beta_0)$, and $\beta_0=\sqrt{|(t-\gamma)/(t+\gamma)|}$\cite{liu2019second}. With this substitution, the corrected Bloch Hamiltonian shows (see the Supplemental Material) agreement with numerical simulations of an open-boundary system, that a finite bulk bandgap exists for all values of the hopping amplitudes except at $t^2 = \gamma^2 \pm \lambda^2$. The zero-gap condition is represented in Fig.~\ref{fig:nhqi}(b) by the solid black lines.

Another important consequence of this extra factor $\beta_0$ is that the bulk spectrum is real-valued for $|t| > |\gamma|$. While there are also edge and corner states, our numerical results show that the entire spectrum is real for arrays of any size whenever $|t| > |\gamma|$. This fact can be related to the pseudo-Hermiticity of the Hamiltonian \cite{liu2019second}.

\begin{figure}
\includegraphics[width=\linewidth]{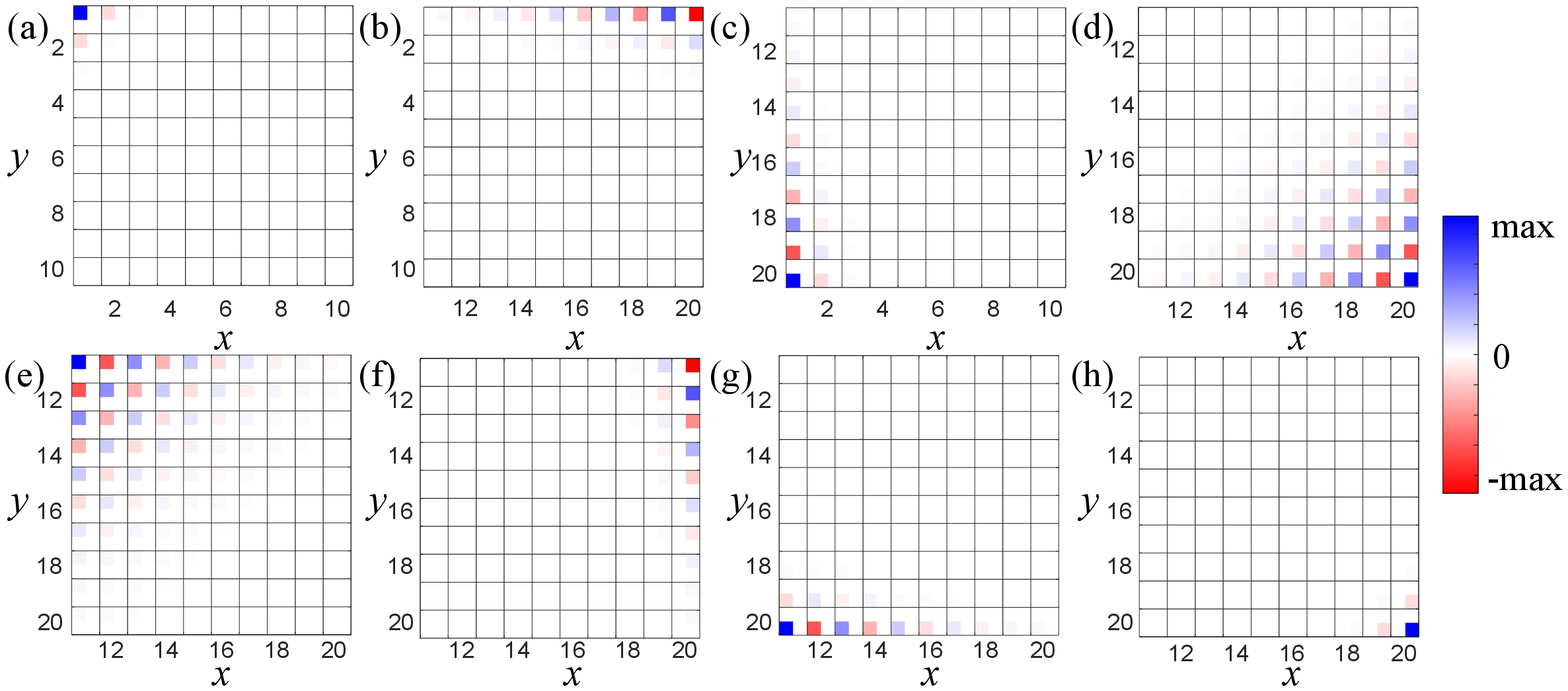}
\caption{\label{fig:sp1} Field distribution of (a-d) four zero-energy right eigenstates and (e-h) corresponding left eigenstates of the non-Hermitian QI with open boundary condition in the ``near-Hermitian" regime $t=0.6, \gamma=0.4, \lambda=1.5$. Domain size: $20 \times 20$ unit cells.}
\end{figure}

Having established the bulk properties of non-Hermitian QIs, we now proceed with investigating the properties of zero-energy corner states supported by a large ($N\times N$ array, $N\gg1$) non-Hermitian QI with open boundary conditions, using Jordan decomposition. 
When the inter-cell hopping strength dominates over the intra-cell one, i.e. $|\lambda| > |t| + |\gamma|$, the system has zero as an eigenvalue of algebraic multiplicity four. The Schur form of this invariant subspace turns out to be approximately a diagonal matrix $J_0=0_{4\times4}$. Thus the left and right basis states in this case are all left/right eigenstates, and are shown in Fig.~\ref{fig:sp1}. The right eigenstates are just the four corner states identified in the Hermitian QI~\cite{benalcazar2017quantized}, albeit with modified spatial localization lengths:
\begin{subequations}\label{eq:nh}
\begin{align}
    &|\psi_1\rangle=\sum_{x,y}(-\frac{t-\gamma}{\lambda})^{x+y}|x,y,1\rangle,\label{eq:supra}\\
    &|\psi_2\rangle=\sum_{x,y}(-\frac{t+\gamma}{\lambda})^{-x}(-\frac{t-\gamma}{\lambda})^y|x,y,2\rangle,\\
    &|\psi_3\rangle=\sum_{x,y}(-\frac{t-\gamma}{\lambda})^{x}(-\frac{t+\gamma}{\lambda})^{-y}|x,y,3\rangle,\\
    &|\psi_4\rangle=\sum_{x,y}(-\frac{t+\gamma}{\lambda})^{-x-y}|x,y,4\rangle.
\end{align}
\end{subequations}
We verify $H\psi_i\approx0$ in the thermodynamic limit $N\to\infty$ in the Supplemental Material.
Just as in the Hermitian QI, each corner hosts one corner state, which has support on only one sublattice. The corresponding left and right eigenstates occupy the same corner. We refer to this parameter regime as ``near-Hermitian". 
All four (right) eigenstates can be resonantly excited using an external source in a way similar to the Hermitian case (see the Supplemental Material). Ref.~\cite{liu2019second} was only able to identify one of the four approximate eigenstates using eigendecomposition.

\begin{figure}
\includegraphics[width=\linewidth]{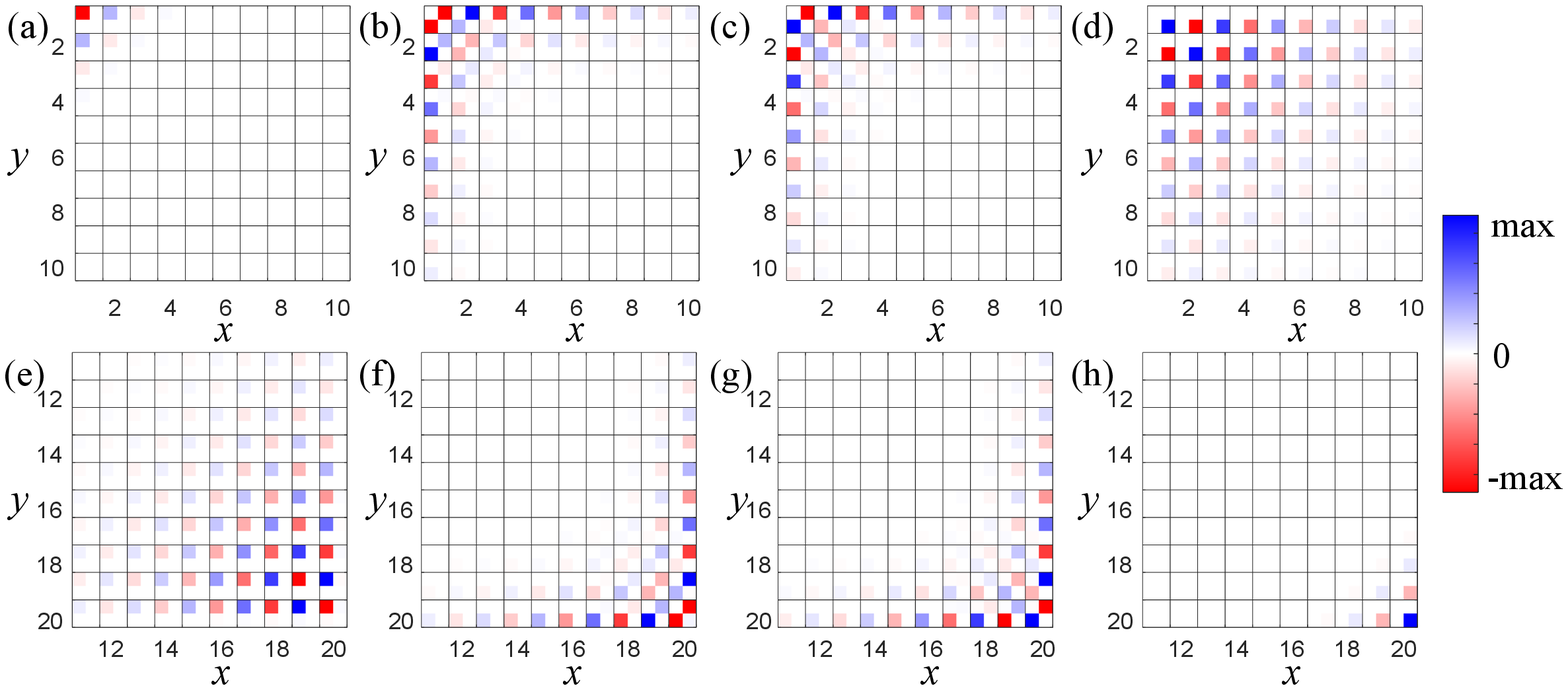}
\caption{\label{fig:sp2} Field distribution of (a-d) four zero-energy right basis states and (e-h) corresponding left basis states of the non-Hermitian QI with open boundary condition in the intermediate regime $t=0.6, \gamma=0.4, \lambda=0.7$. Domain size: $20 \times 20$ unit cells.}
\end{figure}

The situation is very different as we enter the \textit{intermediate} regime $\sqrt{|t^2-\gamma^2|} <|\lambda| < |t|+|\gamma|$ (Fig.~\ref{fig:nhqi}(b) cyan). The system still has zero as an eigenvalue of algebraic multiplicity four, but the Schur form of this invariant subspace is now 
\begin{equation}\label{eq:J0schur}
    J_0=
    \begin{pmatrix}
    0 & -\kappa & 0 & 0 \\
    0 & 0 & 0 & 0 \\
    0 & 0 & 0 & \kappa  \\
    0 & 0 & 0 & 0
 \end{pmatrix},
\end{equation}
where $\kappa$ is a function of parameters. For example, $\kappa\approx0.8246$ when $t=0.6, \gamma=0.4, \lambda=0.7$. The left and right basis states are shown in Fig.~\ref{fig:sp2}. From the form of $J_0$, we know that in this regime the zero energy block is {\it defective}, and only the first (Fig.~\ref{fig:sp2}(a)) and third (Fig.~\ref{fig:sp2}(c)) right basis states are (right) eigenstates; the other two are sometimes called generalized eigenstates. Fig.~\ref{fig:sp2}(a) is similar to Fig.~\ref{fig:sp1}(a), but Fig.~\ref{fig:sp2}(c) is new, and has support on two ($2$ and $3$) sublattices \cite{*[{Similar corner states have been predicted in a Hermitian model in }] [{, but there the bulk is gapless.}] PhysRevB.98.205422}. We refer to the former as ``mono-sublattice", and the latter ``multi-sublattice".
Another major difference from the near-Hermitian regime is that all four right basis states are localized at the top-left corner, while all four left basis states are localized at the bottom-right corner.

We note that this is a perfect example of how a few-level non-Hermitian (or more specifically, defective) model, like Eq.~(\ref{eq:J0schur}), can emerge from a large, quasi-continuous system. Many studies in the field of non-Hermitian physics starts from a two-level model \cite{chen2020revealing,PhysRevLett.118.093002,el2018non}, but our work shows how the collective behavior of lattice sites can serve as an ingredient of these abstract models.

In a perfect lattice like this, an approximate analytical expression of the basis states can be obtained as
\begin{align}
    &|\eta_1^R\rangle=\sum_{x,y}r_1^{x+y}|x,y,1\rangle,\nonumber\\
    &|\eta_2^R\rangle=\sum_{x,y}(r_1^x-r_2^x)r_1^y(|x,y,2\rangle+|y,x,3\rangle),\nonumber\\
    &|\eta_3^R\rangle=\sum_{x,y}(r_1^x-r_2^x)r_1^y(|x,y,2\rangle-|y,x,3\rangle),\nonumber\\
    &|\eta_4^R\rangle=\sum_{x,y}(r_1^x-r_2^x)(r_1^y-r_2^y)|x,y,4\rangle,\nonumber\\
    &\langle{\eta}_1^L|=A_1\sum_{x,y}(r_1^{\bar{x}}-r_2^{\bar{x}})(r_1^{\bar{y}}-r_2^{\bar{y}})\langle x,y,1|,\nonumber\\
    &\langle{\eta}_2^L|=A_2\sum_{x,y}r_1^{\bar{x}}(r_1^{\bar{y}}-r_2^{\bar{y}})(\langle x,y,2|+\langle y,x,3|),\nonumber\\
    &\langle{\eta}_3^L|=A_3\sum_{x,y}r_1^{\bar{x}}(r_1^{\bar{y}}-r_2^{\bar{y}})(\langle x,y,2|-\langle y,x,3|),\nonumber\\
    &\langle{\eta}_4^L|=A_4\sum_{x,y}r_1^{\bar{x}+\bar{y}}\langle x,y,4|,
\end{align}
where $r_1=-(t-\gamma)/\lambda,r_2=-\lambda/(t+\gamma),\bar{x}=N+1-x,\bar{y}=N+1-y$. It can be directly verified that $\langle{\eta}_m^L|{\eta}_n^R\rangle=0$ for $m\neq n$ as required. Normalization constants $A_n\sim r_2^{-2N}$ so that $\langle{\eta}_n^L|{\eta}_n^R\rangle=1$. The normalization constants are huge simply because left and right states are localized at opposite corners. Under this choice
\begin{equation}\label{eq:J0}
    J_0=
    \begin{pmatrix}
    0 & 2\kappa & 0 & 0 \\
    0 & 0 & 0 & 0 \\
    0  & 0  & 0 & \kappa  \\
    0 & 0 & 0 & 0
 \end{pmatrix},
\end{equation}
where $\kappa=t+\gamma-\lambda^2/(t-\gamma)$. $J_0$ and the value of $\kappa$ are slightly different from those obtained from numerical Schur decomposition, purely due to normalization. In a more complicated lattice or a lattice with random variation (e.g., due to limit of fabrication accuracy), it may be impossible to obtain an analytical solution, but we can still rely on the numerical Schur decomposition.

When the inter-cell hopping amplitude is further reduced to $|\lambda| < \sqrt{|t^2-\gamma^2|}$, zero-energy corner states disappear (trivial regime). The three regimes of a square finite-sized non-Hermitian QI with open boundary conditions are summarized by a phase diagram  shown in Fig.~\ref{fig:nhqi}(b). 

\begin{figure}
\includegraphics[width=\linewidth]{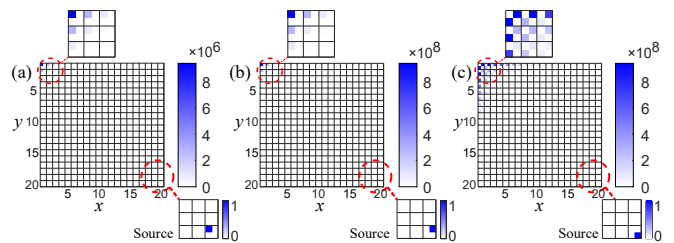}
\caption{\label{fig:resp} The response of a non-Hermitian QI in the intermediate regime to external sources placed at different sub-lattice sites in the lower-right corner of the domain. The source sublattice sites are $1$ (left), $2$ (middle), and $4$ (right). Color: magnitude of the complex field $\psi$. Mono-sublattice (left and middle) and multi-sublattice (right) corner states are predominantly excited (cf. Fig.~\ref{fig:sp2}). Source frequency: $E=0$, small uniform on-site loss $\Gamma=0.01$. Other lattice parameters (domain size and hopping amplitudes) of the tight-binding model: same as in Fig.~\ref{fig:sp2}.}
\end{figure}

 The resonant excitation of our non-Hermitian QI in the intermediate regime is drastically different from that of Hermitian QI. First, we present the results of driven simulations with localized sources, and then interpret the results using the Jordan decomposition that we just obtained. 
 The responses of the system introduced in Fig.~\ref{fig:sp2} (see the caption for the lattice parameters) to external sources (with zero frequency) localized at different sub-lattice sites are shown in Fig.~\ref{fig:resp}(a-c). Surprisingly, our simulations reveal that the response is ``non-local": placing the source at the bottom-right corner gives the strongest excitation of the top-left corner states, whereas in Hermitian systems, one always finds it most efficient to place the source in close proximity of the targeted state's maximum. Moreover, we find that the response is very sensitive to which sublattice the source is on: the mono-sublattice state is predominantly excited by placing the source on the sublattices $1$, $2$, or $3$. On the other hand, the multi-sublattice state is predominantly excited when the source is on the sublattice $4$. Finally, the response in the intermediate regime is much larger compared to that in the near-Hermitian regime (compare Fig.~S3 and Fig.~\ref{fig:resp}).

If we did not have the knowledge of Jordan decomposition of our system, these excitation behaviors would indeed seem very exotic and hard to explain. 

As mentioned earlier, the driven equation $(E-H+i\Gamma)\psi=\xi$ reduces to $(E-J+i\Gamma)\psi'=\xi'$ where $\xi'=P^{-1}\xi,\phi=P\psi'$. To understand the behavior of $H$ near $E=0$, we only need to work in the above mentioned four-dimensional subspace because only the vectors in this subspace can diverge as $1/E$ or faster. Therefore, below we appropriate the notations $\xi'$ and $\psi'$ to just represent the four dimensional vectors. As mentioned, the resolvent $(E-J_0+i\Gamma)^{-1}$ is easy to compute (using Eq.~(\ref{eq:J0})):
\begin{equation}
    (E-J_0)^{-1}=\begin{pmatrix}
    1/E & 2\kappa/E^2 & 0 & 0 \\
    0 & 1/E& 0 & 0 \\
    0  & 0  & 1/E & \kappa/E^2  \\
    0 & 0 & 0 & 1/E
 \end{pmatrix},
\end{equation}
with substitution $E\to E+i\Gamma$.
To calculate $\xi'$ we need to calculate the inner products $\langle\eta_n^L|\xi\rangle$. From the form of ${\eta}_n^L$ we immediately see that placing a source on sublattice $1$ gives $\xi'\propto(1,0,0,0)^T$. By calculating $\psi'=(E-J_0+i\Gamma)^{-1}\xi'$ we know that the mono-sublattice state is excited. Likewise, placing a source on sublattice $4$ induces $\xi'\propto(0,0,0,1)^T$, so the multi-sublattice state is excited. Note that placing a source on either sublattice $2$ or $3$ induces $\xi'\propto(0,1,\pm1,0)^T$, but the mono-sublattice state still dominates due to its faster divergence rate $1/E^2$. This is clearly observed in Fig.~\ref{fig:resp}, where the response to the sources placed on sublattices $2$ and $4$ (middle and right figures) is stronger than that to the source placed on sublattice $1$ (left figure).
Placing the source as far away as possible from the corner states leads to stronger excitation of the latter because this maximizes the inner product with ${\eta}_n^L$. The huge amplitude of the response $\psi$ is mainly due to the exponentially large normalization constants $A_n$.

We demonstrated the issues of using eigendecomposition in non-Hermitian lattice systems and proposed using Jordan decomposition as a replacement. Jordan decomposition is a reasonable choice since it is a generalization of eigendecomposition and because of its close connection to the equation of motion of the system. We provide procedures of how it can be achieved numerically in a system with spectrally isolated modes. We used a non-Hermitian quadrupole insulator with asymmetric intracell coupling strengths as an example to show how a few-level defective Hamiltonian can emerge from a large lattice and that the delicate and exotic response of the system to external drives can be beautifully explained by the Jordan basis states and the Jordan form. 

\begin{acknowledgements}
This work was supported by the Office of Naval Research (ONR) under Grant No. N00014-17-1-2161, by the National Science Foundation (NSF) under Grant No. DMR-1741788, and by the Cornell Center for Materials Research with funding from the NSF MRSEC program (DMR-1719875). M. J. was supported in part by Kwanjeong Educational Foundation.
\end{acknowledgements}
\bibliography{nhqi}

\providecommand{\noopsort}[1]{}\providecommand{\singleletter}[1]{#1}%
\begin{thebibliography}{27}%
\makeatletter
\providecommand \@ifxundefined [1]{%
 \@ifx{#1\undefined}
}%
\providecommand \@ifnum [1]{%
 \ifnum #1\expandafter \@firstoftwo
 \else \expandafter \@secondoftwo
 \fi
}%
\providecommand \@ifx [1]{%
 \ifx #1\expandafter \@firstoftwo
 \else \expandafter \@secondoftwo
 \fi
}%
\providecommand \natexlab [1]{#1}%
\providecommand \enquote  [1]{``#1''}%
\providecommand \bibnamefont  [1]{#1}%
\providecommand \bibfnamefont [1]{#1}%
\providecommand \citenamefont [1]{#1}%
\providecommand \href@noop [0]{\@secondoftwo}%
\providecommand \href [0]{\begingroup \@sanitize@url \@href}%
\providecommand \@href[1]{\@@startlink{#1}\@@href}%
\providecommand \@@href[1]{\endgroup#1\@@endlink}%
\providecommand \@sanitize@url [0]{\catcode `\\12\catcode `\$12\catcode
  `\&12\catcode `\#12\catcode `\^12\catcode `\_12\catcode `\%12\relax}%
\providecommand \@@startlink[1]{}%
\providecommand \@@endlink[0]{}%
\providecommand \url  [0]{\begingroup\@sanitize@url \@url }%
\providecommand \@url [1]{\endgroup\@href {#1}{\urlprefix }}%
\providecommand \urlprefix  [0]{URL }%
\providecommand \Eprint [0]{\href }%
\providecommand \doibase [0]{http://dx.doi.org/}%
\providecommand \selectlanguage [0]{\@gobble}%
\providecommand \bibinfo  [0]{\@secondoftwo}%
\providecommand \bibfield  [0]{\@secondoftwo}%
\providecommand \translation [1]{[#1]}%
\providecommand \BibitemOpen [0]{}%
\providecommand \bibitemStop [0]{}%
\providecommand \bibitemNoStop [0]{.\EOS\space}%
\providecommand \EOS [0]{\spacefactor3000\relax}%
\providecommand \BibitemShut  [1]{\csname bibitem#1\endcsname}%
\let\auto@bib@innerbib\@empty
\bibitem [{\citenamefont {Baker}(1984)}]{baker_pra84}%
  \BibitemOpen
  \bibfield  {author} {\bibinfo {author} {\bibfnamefont {H.~C.}\ \bibnamefont
  {Baker}},\ }\href@noop {} {\bibfield  {journal} {\bibinfo  {journal} {Phys.
  Rev. A}\ }\textbf {\bibinfo {volume} {30}},\ \bibinfo {pages} {773} (\bibinfo
  {year} {1984})}\BibitemShut {NoStop}%
\bibitem [{\citenamefont {Lopata}\ and\ \citenamefont
  {Govind}(2013)}]{lopata_jctc13}%
  \BibitemOpen
  \bibfield  {author} {\bibinfo {author} {\bibfnamefont {K.}~\bibnamefont
  {Lopata}}\ and\ \bibinfo {author} {\bibfnamefont {N.}~\bibnamefont
  {Govind}},\ }\href@noop {} {\bibfield  {journal} {\bibinfo  {journal} {J.
  Chem. Th. and Comp.}\ }\textbf {\bibinfo {volume} {9}},\ \bibinfo {pages}
  {4939} (\bibinfo {year} {2013})}\BibitemShut {NoStop}%
\bibitem [{\citenamefont {Heiss}(2004)}]{heiss2004exceptional}%
  \BibitemOpen
  \bibfield  {author} {\bibinfo {author} {\bibfnamefont {W.}~\bibnamefont
  {Heiss}},\ }\href@noop {} {\bibfield  {journal} {\bibinfo  {journal} {J.
  Phys. Math. Gen.}\ }\textbf {\bibinfo {volume} {37}},\ \bibinfo {pages}
  {2455} (\bibinfo {year} {2004})}\BibitemShut {NoStop}%
\bibitem [{\citenamefont {Berry}(2004)}]{berry2004physics}%
  \BibitemOpen
  \bibfield  {author} {\bibinfo {author} {\bibfnamefont {M.~V.}\ \bibnamefont
  {Berry}},\ }\href@noop {} {\bibfield  {journal} {\bibinfo  {journal}
  {Czechoslov. J. Phys.}\ }\textbf {\bibinfo {volume} {54}},\ \bibinfo {pages}
  {1039} (\bibinfo {year} {2004})}\BibitemShut {NoStop}%
\bibitem [{\citenamefont {Moiseyev}(2011)}]{moiseyev2011non}%
  \BibitemOpen
  \bibfield  {author} {\bibinfo {author} {\bibfnamefont {N.}~\bibnamefont
  {Moiseyev}},\ }\href@noop {} {\emph {\bibinfo {title} {Non-Hermitian quantum
  mechanics}}}\ (\bibinfo  {publisher} {Cambridge University Press},\ \bibinfo
  {year} {2011})\BibitemShut {NoStop}%
\bibitem [{\citenamefont {Feng}\ \emph {et~al.}(2017)\citenamefont {Feng},
  \citenamefont {El-Ganainy},\ and\ \citenamefont {Ge}}]{liang_feng_nphot17}%
  \BibitemOpen
  \bibfield  {author} {\bibinfo {author} {\bibfnamefont {L.}~\bibnamefont
  {Feng}}, \bibinfo {author} {\bibfnamefont {R.}~\bibnamefont {El-Ganainy}}, \
  and\ \bibinfo {author} {\bibfnamefont {L.}~\bibnamefont {Ge}},\ }\href@noop
  {} {\bibfield  {journal} {\bibinfo  {journal} {Nat. Phot.}\ }\textbf
  {\bibinfo {volume} {11}},\ \bibinfo {pages} {752} (\bibinfo {year}
  {2017})}\BibitemShut {NoStop}%
\bibitem [{\citenamefont {El-Ganainy}\ \emph {et~al.}(2018)\citenamefont
  {El-Ganainy}, \citenamefont {Makris}, \citenamefont {Khajavikhan},
  \citenamefont {Musslimani}, \citenamefont {Rotter},\ and\ \citenamefont
  {Christodoulides}}]{el2018non}%
  \BibitemOpen
  \bibfield  {author} {\bibinfo {author} {\bibfnamefont {R.}~\bibnamefont
  {El-Ganainy}}, \bibinfo {author} {\bibfnamefont {K.~G.}\ \bibnamefont
  {Makris}}, \bibinfo {author} {\bibfnamefont {M.}~\bibnamefont {Khajavikhan}},
  \bibinfo {author} {\bibfnamefont {Z.~H.}\ \bibnamefont {Musslimani}},
  \bibinfo {author} {\bibfnamefont {S.}~\bibnamefont {Rotter}}, \ and\ \bibinfo
  {author} {\bibfnamefont {D.~N.}\ \bibnamefont {Christodoulides}},\
  }\href@noop {} {\bibfield  {journal} {\bibinfo  {journal} {Nat. Phys.}\
  }\textbf {\bibinfo {volume} {14}},\ \bibinfo {pages} {11} (\bibinfo {year}
  {2018})}\BibitemShut {NoStop}%
\bibitem [{\citenamefont {Golub}\ and\ \citenamefont
  {Van~Loan}(2013)}]{golub2013matrix}%
  \BibitemOpen
  \bibfield  {author} {\bibinfo {author} {\bibfnamefont {G.~H.}\ \bibnamefont
  {Golub}}\ and\ \bibinfo {author} {\bibfnamefont {C.~F.}\ \bibnamefont
  {Van~Loan}},\ }\href@noop {} {\emph {\bibinfo {title} {Matrix
  computations}}},\ \bibinfo {edition} {4th}\ ed.\ (\bibinfo  {publisher}
  {Johns Hopkins University Press},\ \bibinfo {year} {2013})\BibitemShut
  {NoStop}%
\bibitem [{\citenamefont {Lee}(2016)}]{lee2016anomalous}%
  \BibitemOpen
  \bibfield  {author} {\bibinfo {author} {\bibfnamefont {T.~E.}\ \bibnamefont
  {Lee}},\ }\href@noop {} {\bibfield  {journal} {\bibinfo  {journal} {Phys.
  Rev. Lett.}\ }\textbf {\bibinfo {volume} {116}},\ \bibinfo {pages} {133903}
  (\bibinfo {year} {2016})}\BibitemShut {NoStop}%
\bibitem [{\citenamefont {Bender}(2007)}]{bender2007making}%
  \BibitemOpen
  \bibfield  {author} {\bibinfo {author} {\bibfnamefont {C.~M.}\ \bibnamefont
  {Bender}},\ }\href@noop {} {\bibfield  {journal} {\bibinfo  {journal} {Rep.
  Prog. Phys.}\ }\textbf {\bibinfo {volume} {70}},\ \bibinfo {pages} {947}
  (\bibinfo {year} {2007})}\BibitemShut {NoStop}%
\bibitem [{\citenamefont {West}\ \emph {et~al.}(2010)\citenamefont {West},
  \citenamefont {Kottos},\ and\ \citenamefont
  {Prosen}}]{PhysRevLett.104.054102}%
  \BibitemOpen
  \bibfield  {author} {\bibinfo {author} {\bibfnamefont {C.~T.}\ \bibnamefont
  {West}}, \bibinfo {author} {\bibfnamefont {T.}~\bibnamefont {Kottos}}, \ and\
  \bibinfo {author} {\bibfnamefont {T.}~\bibnamefont {Prosen}},\ }\href
  {\doibase 10.1103/PhysRevLett.104.054102} {\bibfield  {journal} {\bibinfo
  {journal} {Phys. Rev. Lett.}\ }\textbf {\bibinfo {volume} {104}},\ \bibinfo
  {pages} {054102} (\bibinfo {year} {2010})}\BibitemShut {NoStop}%
\bibitem [{\citenamefont {Bender}\ \emph {et~al.}(2013)\citenamefont {Bender},
  \citenamefont {Gianfreda}, \citenamefont {\"Ozdemir}, \citenamefont {Peng},\
  and\ \citenamefont {Yang}}]{PhysRevA.88.062111}%
  \BibitemOpen
  \bibfield  {author} {\bibinfo {author} {\bibfnamefont {C.~M.}\ \bibnamefont
  {Bender}}, \bibinfo {author} {\bibfnamefont {M.}~\bibnamefont {Gianfreda}},
  \bibinfo {author} {\bibfnamefont {{\c{S}}.~K.}\ \bibnamefont {\"Ozdemir}},
  \bibinfo {author} {\bibfnamefont {B.}~\bibnamefont {Peng}}, \ and\ \bibinfo
  {author} {\bibfnamefont {L.}~\bibnamefont {Yang}},\ }\href {\doibase
  10.1103/PhysRevA.88.062111} {\bibfield  {journal} {\bibinfo  {journal} {Phys.
  Rev. A}\ }\textbf {\bibinfo {volume} {88}},\ \bibinfo {pages} {062111}
  (\bibinfo {year} {2013})}\BibitemShut {NoStop}%
\bibitem [{\citenamefont {Mostafazadeh}(2002)}]{mostafazadeh2002pseudo}%
  \BibitemOpen
  \bibfield  {author} {\bibinfo {author} {\bibfnamefont {A.}~\bibnamefont
  {Mostafazadeh}},\ }\href@noop {} {\bibfield  {journal} {\bibinfo  {journal}
  {J. Math. Phys.}\ }\textbf {\bibinfo {volume} {43}},\ \bibinfo {pages} {205}
  (\bibinfo {year} {2002})}\BibitemShut {NoStop}%
\bibitem [{\citenamefont {Liu}\ \emph {et~al.}(2019)\citenamefont {Liu},
  \citenamefont {Zhang}, \citenamefont {Ai}, \citenamefont {Gong},
  \citenamefont {Kawabata}, \citenamefont {Ueda},\ and\ \citenamefont
  {Nori}}]{liu2019second}%
  \BibitemOpen
  \bibfield  {author} {\bibinfo {author} {\bibfnamefont {T.}~\bibnamefont
  {Liu}}, \bibinfo {author} {\bibfnamefont {Y.-R.}\ \bibnamefont {Zhang}},
  \bibinfo {author} {\bibfnamefont {Q.}~\bibnamefont {Ai}}, \bibinfo {author}
  {\bibfnamefont {Z.}~\bibnamefont {Gong}}, \bibinfo {author} {\bibfnamefont
  {K.}~\bibnamefont {Kawabata}}, \bibinfo {author} {\bibfnamefont
  {M.}~\bibnamefont {Ueda}}, \ and\ \bibinfo {author} {\bibfnamefont
  {F.}~\bibnamefont {Nori}},\ }\href@noop {} {\bibfield  {journal} {\bibinfo
  {journal} {Phys. Rev. Lett.}\ }\textbf {\bibinfo {volume} {122}},\ \bibinfo
  {pages} {076801} (\bibinfo {year} {2019})}\BibitemShut {NoStop}%
\bibitem [{\citenamefont {Kunst}\ \emph {et~al.}(2018)\citenamefont {Kunst},
  \citenamefont {Edvardsson}, \citenamefont {Budich},\ and\ \citenamefont
  {Bergholtz}}]{PhysRevLett.121.026808}%
  \BibitemOpen
  \bibfield  {author} {\bibinfo {author} {\bibfnamefont {F.~K.}\ \bibnamefont
  {Kunst}}, \bibinfo {author} {\bibfnamefont {E.}~\bibnamefont {Edvardsson}},
  \bibinfo {author} {\bibfnamefont {J.~C.}\ \bibnamefont {Budich}}, \ and\
  \bibinfo {author} {\bibfnamefont {E.~J.}\ \bibnamefont {Bergholtz}},\ }\href
  {\doibase 10.1103/PhysRevLett.121.026808} {\bibfield  {journal} {\bibinfo
  {journal} {Phys. Rev. Lett.}\ }\textbf {\bibinfo {volume} {121}},\ \bibinfo
  {pages} {026808} (\bibinfo {year} {2018})}\BibitemShut {NoStop}%
\bibitem [{\citenamefont {Yao}\ and\ \citenamefont {Wang}(2018)}]{yao2018edge}%
  \BibitemOpen
  \bibfield  {author} {\bibinfo {author} {\bibfnamefont {S.}~\bibnamefont
  {Yao}}\ and\ \bibinfo {author} {\bibfnamefont {Z.}~\bibnamefont {Wang}},\
  }\href@noop {} {\bibfield  {journal} {\bibinfo  {journal} {Phys. Rev. Lett.}\
  }\textbf {\bibinfo {volume} {121}},\ \bibinfo {pages} {086803} (\bibinfo
  {year} {2018})}\BibitemShut {NoStop}%
\bibitem [{\citenamefont {Kawabata}\ \emph {et~al.}(2018)\citenamefont
  {Kawabata}, \citenamefont {Shiozaki},\ and\ \citenamefont
  {Ueda}}]{PhysRevB.98.165148}%
  \BibitemOpen
  \bibfield  {author} {\bibinfo {author} {\bibfnamefont {K.}~\bibnamefont
  {Kawabata}}, \bibinfo {author} {\bibfnamefont {K.}~\bibnamefont {Shiozaki}},
  \ and\ \bibinfo {author} {\bibfnamefont {M.}~\bibnamefont {Ueda}},\ }\href
  {\doibase 10.1103/PhysRevB.98.165148} {\bibfield  {journal} {\bibinfo
  {journal} {Phys. Rev. B}\ }\textbf {\bibinfo {volume} {98}},\ \bibinfo
  {pages} {165148} (\bibinfo {year} {2018})}\BibitemShut {NoStop}%
\bibitem [{\citenamefont {Longhi}\ \emph {et~al.}(2015)\citenamefont {Longhi},
  \citenamefont {Gatti},\ and\ \citenamefont {Della~Valle}}]{longhi2015robust}%
  \BibitemOpen
  \bibfield  {author} {\bibinfo {author} {\bibfnamefont {S.}~\bibnamefont
  {Longhi}}, \bibinfo {author} {\bibfnamefont {D.}~\bibnamefont {Gatti}}, \
  and\ \bibinfo {author} {\bibfnamefont {G.}~\bibnamefont {Della~Valle}},\
  }\href@noop {} {\bibfield  {journal} {\bibinfo  {journal} {Phys. Rev. B}\
  }\textbf {\bibinfo {volume} {92}},\ \bibinfo {pages} {094204} (\bibinfo
  {year} {2015})}\BibitemShut {NoStop}%
\bibitem [{\citenamefont {McDonald}\ \emph {et~al.}(2018)\citenamefont
  {McDonald}, \citenamefont {Pereg-Barnea},\ and\ \citenamefont
  {Clerk}}]{PhysRevX.8.041031}%
  \BibitemOpen
  \bibfield  {author} {\bibinfo {author} {\bibfnamefont {A.}~\bibnamefont
  {McDonald}}, \bibinfo {author} {\bibfnamefont {T.}~\bibnamefont
  {Pereg-Barnea}}, \ and\ \bibinfo {author} {\bibfnamefont {A.~A.}\
  \bibnamefont {Clerk}},\ }\href {\doibase 10.1103/PhysRevX.8.041031}
  {\bibfield  {journal} {\bibinfo  {journal} {Phys. Rev. X}\ }\textbf {\bibinfo
  {volume} {8}},\ \bibinfo {pages} {041031} (\bibinfo {year}
  {2018})}\BibitemShut {NoStop}%
\bibitem [{\citenamefont {Schomerus}(2020)}]{PhysRevResearch.2.013058}%
  \BibitemOpen
  \bibfield  {author} {\bibinfo {author} {\bibfnamefont {H.}~\bibnamefont
  {Schomerus}},\ }\href {\doibase 10.1103/PhysRevResearch.2.013058} {\bibfield
  {journal} {\bibinfo  {journal} {Phys. Rev. Research}\ }\textbf {\bibinfo
  {volume} {2}},\ \bibinfo {pages} {013058} (\bibinfo {year}
  {2020})}\BibitemShut {NoStop}%
\bibitem [{\citenamefont {Jin}\ and\ \citenamefont
  {Song}(2019)}]{PhysRevB.99.081103}%
  \BibitemOpen
  \bibfield  {author} {\bibinfo {author} {\bibfnamefont {L.}~\bibnamefont
  {Jin}}\ and\ \bibinfo {author} {\bibfnamefont {Z.}~\bibnamefont {Song}},\
  }\href {\doibase 10.1103/PhysRevB.99.081103} {\bibfield  {journal} {\bibinfo
  {journal} {Phys. Rev. B}\ }\textbf {\bibinfo {volume} {99}},\ \bibinfo
  {pages} {081103} (\bibinfo {year} {2019})}\BibitemShut {NoStop}%
\bibitem [{\citenamefont {Chen}\ \emph {et~al.}(2020)\citenamefont {Chen},
  \citenamefont {Liu}, \citenamefont {Luan}, \citenamefont {Liu}, \citenamefont
  {Wang}, \citenamefont {Zhu}, \citenamefont {Li}, \citenamefont {Gu},
  \citenamefont {Liang}, \citenamefont {Gao} \emph
  {et~al.}}]{chen2020revealing}%
  \BibitemOpen
  \bibfield  {author} {\bibinfo {author} {\bibfnamefont {H.-Z.}\ \bibnamefont
  {Chen}}, \bibinfo {author} {\bibfnamefont {T.}~\bibnamefont {Liu}}, \bibinfo
  {author} {\bibfnamefont {H.-Y.}\ \bibnamefont {Luan}}, \bibinfo {author}
  {\bibfnamefont {R.-J.}\ \bibnamefont {Liu}}, \bibinfo {author} {\bibfnamefont
  {X.-Y.}\ \bibnamefont {Wang}}, \bibinfo {author} {\bibfnamefont {X.-F.}\
  \bibnamefont {Zhu}}, \bibinfo {author} {\bibfnamefont {Y.-B.}\ \bibnamefont
  {Li}}, \bibinfo {author} {\bibfnamefont {Z.-M.}\ \bibnamefont {Gu}}, \bibinfo
  {author} {\bibfnamefont {S.-J.}\ \bibnamefont {Liang}}, \bibinfo {author}
  {\bibfnamefont {H.}~\bibnamefont {Gao}},  \emph {et~al.},\ }\href@noop {}
  {\bibfield  {journal} {\bibinfo  {journal} {Nature Physics}\ }\textbf
  {\bibinfo {volume} {16}},\ \bibinfo {pages} {571} (\bibinfo {year}
  {2020})}\BibitemShut {NoStop}%
\bibitem [{\citenamefont {Benalcazar}\ \emph {et~al.}(2017)\citenamefont
  {Benalcazar}, \citenamefont {Bernevig},\ and\ \citenamefont
  {Hughes}}]{benalcazar2017quantized}%
  \BibitemOpen
  \bibfield  {author} {\bibinfo {author} {\bibfnamefont {W.~A.}\ \bibnamefont
  {Benalcazar}}, \bibinfo {author} {\bibfnamefont {B.~A.}\ \bibnamefont
  {Bernevig}}, \ and\ \bibinfo {author} {\bibfnamefont {T.~L.}\ \bibnamefont
  {Hughes}},\ }\href@noop {} {\bibfield  {journal} {\bibinfo  {journal}
  {Science}\ }\textbf {\bibinfo {volume} {357}},\ \bibinfo {pages} {61}
  (\bibinfo {year} {2017})}\BibitemShut {NoStop}%
\bibitem [{\citenamefont {Lieu}(2018)}]{lieu2018topological}%
  \BibitemOpen
  \bibfield  {author} {\bibinfo {author} {\bibfnamefont {S.}~\bibnamefont
  {Lieu}},\ }\href@noop {} {\bibfield  {journal} {\bibinfo  {journal} {Phys.
  Rev. B}\ }\textbf {\bibinfo {volume} {97}},\ \bibinfo {pages} {045106}
  (\bibinfo {year} {2018})}\BibitemShut {NoStop}%
\bibitem [{\citenamefont {Yin}\ \emph {et~al.}(2018)\citenamefont {Yin},
  \citenamefont {Jiang}, \citenamefont {Li}, \citenamefont {L{\"u}},\ and\
  \citenamefont {Chen}}]{yin2018geometrical}%
  \BibitemOpen
  \bibfield  {author} {\bibinfo {author} {\bibfnamefont {C.}~\bibnamefont
  {Yin}}, \bibinfo {author} {\bibfnamefont {H.}~\bibnamefont {Jiang}}, \bibinfo
  {author} {\bibfnamefont {L.}~\bibnamefont {Li}}, \bibinfo {author}
  {\bibfnamefont {R.}~\bibnamefont {L{\"u}}}, \ and\ \bibinfo {author}
  {\bibfnamefont {S.}~\bibnamefont {Chen}},\ }\href@noop {} {\bibfield
  {journal} {\bibinfo  {journal} {Phys. Rev. A}\ }\textbf {\bibinfo {volume}
  {97}},\ \bibinfo {pages} {052115} (\bibinfo {year} {2018})}\BibitemShut
  {NoStop}%
\bibitem [{\citenamefont {Li}\ \emph {et~al.}(2018)\citenamefont {Li},
  \citenamefont {Umer},\ and\ \citenamefont {Gong}}]{PhysRevB.98.205422}%
  \BibitemOpen
  \bibfield  {author} {\bibinfo {author} {\bibfnamefont {L.}~\bibnamefont
  {Li}}, \bibinfo {author} {\bibfnamefont {M.}~\bibnamefont {Umer}}, \ and\
  \bibinfo {author} {\bibfnamefont {J.}~\bibnamefont {Gong}},\ }\href {\doibase
  10.1103/PhysRevB.98.205422} {\bibfield  {journal} {\bibinfo  {journal} {Phys.
  Rev. B}\ }\textbf {\bibinfo {volume} {98}},\ \bibinfo {pages} {205422}
  (\bibinfo {year} {2018})}\BibitemShut {NoStop}%
\bibitem [{\citenamefont {Hassan}\ \emph {et~al.}(2017)\citenamefont {Hassan},
  \citenamefont {Zhen}, \citenamefont {Solja\ifmmode \check{c}\else
  \v{c}\fi{}i\ifmmode~\acute{c}\else \'{c}\fi{}}, \citenamefont {Khajavikhan},\
  and\ \citenamefont {Christodoulides}}]{PhysRevLett.118.093002}%
  \BibitemOpen
  \bibfield  {author} {\bibinfo {author} {\bibfnamefont {A.~U.}\ \bibnamefont
  {Hassan}}, \bibinfo {author} {\bibfnamefont {B.}~\bibnamefont {Zhen}},
  \bibinfo {author} {\bibfnamefont {M.}~\bibnamefont {Solja\ifmmode
  \check{c}\else \v{c}\fi{}i\ifmmode~\acute{c}\else \'{c}\fi{}}}, \bibinfo
  {author} {\bibfnamefont {M.}~\bibnamefont {Khajavikhan}}, \ and\ \bibinfo
  {author} {\bibfnamefont {D.~N.}\ \bibnamefont {Christodoulides}},\ }\href
  {\doibase 10.1103/PhysRevLett.118.093002} {\bibfield  {journal} {\bibinfo
  {journal} {Phys. Rev. Lett.}\ }\textbf {\bibinfo {volume} {118}},\ \bibinfo
  {pages} {093002} (\bibinfo {year} {2017})}\BibitemShut {NoStop}%
\end{thebibliography}%


\providecommand{\noopsort}[1]{}\providecommand{\singleletter}[1]{#1}%
\begin{thebibliography}{1}

\bibitem{asboth2016short}
J.K. Asb{\'o}th, L.~Oroszl{\'a}ny, and A.P. P{\'a}lyi.
\newblock {\em A Short Course on Topological Insulators: Band Structure and
  Edge States in One and Two Dimensions}.
\newblock Lecture Notes in Physics. Springer International Publishing, 2016.

\bibitem{ni2019observation}
Xiang Ni, Matthew Weiner, Andrea Al{\`u}, and Alexander~B Khanikaev.
\newblock Observation of higher-order topological acoustic states protected by
  generalized chiral symmetry.
\newblock {\em Nat. Mat.}, 18(2):113, 2019.

\bibitem{PhysRevB.98.205422}
Linhu Li, Muhammad Umer, and Jiangbin Gong.
\newblock Direct prediction of corner state configurations from edge winding
  numbers in two- and three-dimensional chiral-symmetric lattice systems.
\newblock {\em Phys. Rev. B}, 98:205422, Nov 2018.

\end{thebibliography}

\end{document}


\maketitle
\section{Spectrum of an open-boundary non-Hermitian quadrupole insulator (QI)}
First we briefly show the derivation of the bulk spectrum of an open-boundary non-Hermitian QI, and follwing it, the gap-closing condition and when the bulk spectrum is real.

The Bloch Hamiltonian of the non-Hermitian QI corresponding to Fig. 1(a) is
\begin{equation}
    H(k_x,k_y)=[t+\lambda\cos k_y]\tau_x+[\lambda\sin k_y+i\gamma]\tau_y+[t+\lambda\cos k_x]\tau_z\sigma_x+[\lambda\sin k_x+i\gamma]\tau_z\sigma_y,
\end{equation}
where we have set the lattice constant $a_0=1$, and $\tau_i$ and $\sigma_i (i=x,y,z)$ are Pauli matrices for degrees of freedom within a unit cell.

As mentioned in the paper, to correctly calculate the bulk spectrum of a non-Hermitian QI under open boundary condition, one has to make the following substitution in the Bloch Hamiltonian: $(k_x,k_y)\to(k_x-i\ln \beta_0,k_y-i\ln \beta_0)$, where $\beta_0=\sqrt{|(t-\gamma)/(t+\gamma)|}$.

The eigenvalues $E$ of the resulting Hamiltonian are doubly degenerate and 
\begin{equation}
    E^2(k_x,k_y)=2(t^2-\gamma^2+\lambda^2)+\lambda\sqrt{|t^2-\gamma^2|}[{\rm sgn}(t+\gamma)(e^{ik_x}+e^{ik_y})+{\rm sgn}(t-\gamma)(e^{-ik_x}+e^{-ik_y})].
\end{equation}
It is straightforward to see that the bulk spectrum is real when $|t|>|\gamma|$. $E=0$ (for some real $(k_x,k_y)$) gives the gap-closing condition $t^2=\gamma^2\pm\lambda^2$.

\begin{figure}
\centering
\includegraphics[width=\linewidth]{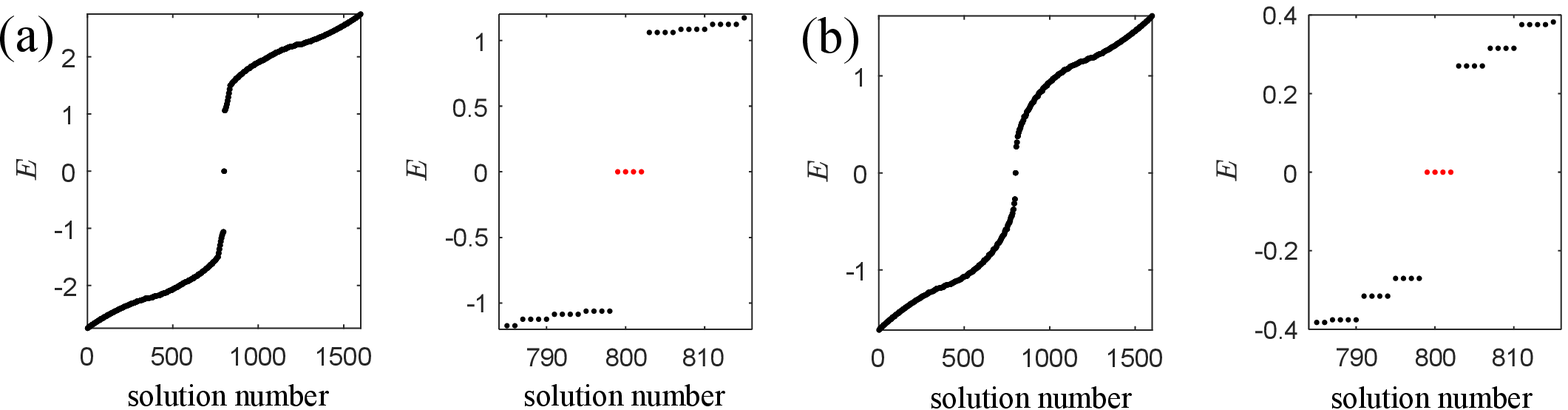}
\caption{Spectra of open-boundary non-Hermitian QIs. (a) Near-Hermitian regime. Left: full spectrum. Right: a zoom-in at the zero-energy corner states (red). Parameters: same as in Fig.~2 ($t=0.6, \gamma=0.4, \lambda=1.5, N=20$). (b) Intermediate regime. Left: full spectrum. Right: a zoom-in at the zero-energy corner states (red). Parameters: same as in Fig.~3 ($t=0.6, \gamma=0.4, \lambda=0.7, N=20$).}
\label{fig:spectrum}
\end{figure}

The complete spectrum of the open-boundary non-Hermitian QI in the ``near-Hermitian" regime, with parameters the same as in Fig.~2 is shown in Fig.~\ref{fig:spectrum}(a). The bulk bandgap as predicted in the previous section, as well as four eigenvalues of $E=0$ (see the zoon-in at right) are clearly visible.

The complete spectrum of the open-boundary non-Hermitian QI in the ``intermediate" regime, with parameters the same as in Fig.~3 is shown in Fig.~\ref{fig:spectrum}(b). The bulk bandgap is smaller but still clear, and four eigenvalues of $E=0$ are present. Note that this only means the algebraic multiplicity of the $E=0$ eigenvalue is 4. As demonstrated in the paper, there are only 2 linearly independent corner states, i.e., the geometric multiplicity of the $E=0$ eigenvalue is $2$.

As mentioned in the paper, both parameter sets fall in to the region where the system has a purely real spectrum.

\section{Analytic verification of corner states}
In this section we verify that the corner states with field distribution given in the paper are indeed approximate eigenstates of the Hamiltonian.

In the near Hermitian regime, there are 4 corner states, with field distribution in the same fashion, see Eq. (3). Similar analytical solutions have been obtained for localized states in Hermitian systems~\cite{asboth2016short,ni2019observation,PhysRevB.98.205422}. Here we just demonstrate $H|\psi_1\rangle\approx0$. Apparently $\langle x,y,1|H|\psi_1\rangle=\langle x,y,4|H|\psi_1\rangle=0$ for all $x,y$.
\begin{align}
    \langle x,y,2|H|\psi_1\rangle=(t-\gamma)(-\frac{t-\gamma}{\lambda})^{x+y}+\lambda(-\frac{t-\gamma}{\lambda})^{x+1+y}=0,\\
    \langle x,y,3|H|\psi_1\rangle=(t-\gamma)(-\frac{t-\gamma}{\lambda})^{x+y}+\lambda(-\frac{t-\gamma}{\lambda})^{x+y+1}=0.
\end{align}
The only exception to the above two equations happens at edges $x=N$ and $y=N$. But in the thermodynamic limit $N\to\infty$ the field is exponentially weak at those edges. Thus $\psi_1$ is an approximate eigenstate of the Hamiltonian, and so is $\psi_2,\psi_3,\psi_4$.

In the intermediate regime, $\psi_1$ survives as the superlocalized state, so we only need to verify that the sublocalized state $\eta_3^R$ in Eq. (5) is an approximate eigenstate. Again it is apparent that $\langle x,y,2|H|\eta_3^R\rangle=\langle x,y,3|H|\eta_3^R\rangle=0$ for all $x,y$.
\begin{align}
    \langle x,y,1|H|\eta_3^R\rangle&=(t+\gamma)(r_1^x-r_2^x)r_1^y+\lambda(r_1^{x-1}-r_2^{x-1})r_1^y-(t+\gamma)(r_1^y-r_2^y)r_1^x-\lambda(r_1^{y-1}-r_2^{y-1})r_1^x\nonumber\\
    &=(t+\gamma)r_1^{x+y}+\lambda r_1^{x-1+y}-(t+\gamma)r_1^{y+x}-\lambda r_1^{y-1+x}=0,\\
    \langle x,y,4|H|\eta_3^R\rangle&=(t-\gamma)(r_1^x-r_2^x)r_1^y+\lambda(r_1^{x}-r_2^{x})r_1^{y+1}+(t-\gamma)(r_1^y-r_2^y)r_1^x+\lambda(r_1^{y}-r_2^{y})r_1^{x+1}\nonumber\\
    &=0+0=0.
\end{align}
Note that the form of the solution satisfies $\eta_3^R(0,y,2)=\eta_3^R(x,0,3)=0$ which is compatible with the open boundary condition, so the above calculation is also valid for boundary sites $x=1$ and $y=1$. Again the only exception happens at edges $x=N$ and $y=N$, but since the field is exponentially weak there, $\eta_3^R$ is an approximate eigenstate of the Hamiltonian.

\section{Additional examples of excitation of corner states in the intermediate regime}
\begin{figure}
\centering
\includegraphics[width=\linewidth]{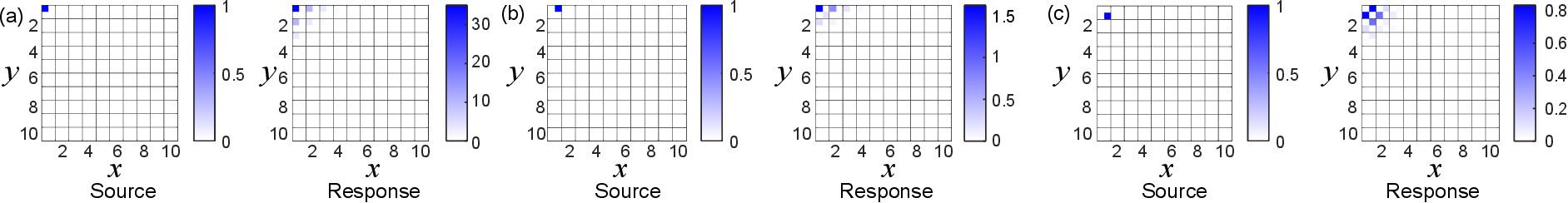}
\caption{The response of a non-Hermitian QI in the intermediate regime to sources placed at different sub-lattice sites in the top-left corner of the domain. The source sublattice sites are $1$ (left), $2$ (middle), and $4$ (right). Color in response plots: magnitude of the complex field $\psi$. Source frequency: $E=0$, uniform on-site loss: $\Gamma=0.01$. Other lattice parameters (domain size and hopping amplitudes) of the tight-binding model: same as in Fig.~3.}
\label{fig:more}
\end{figure}

As mentioned in the main text, in the intermediate regime a source at the top-left corner will only lead to much weaker response compared to a source at the bottom right corner. We show the data here in Fig.~\ref{fig:more}.

\section{Excitation of corner states in the near-Hermitian regime}
\begin{figure}
\centering
\includegraphics[width=\linewidth]{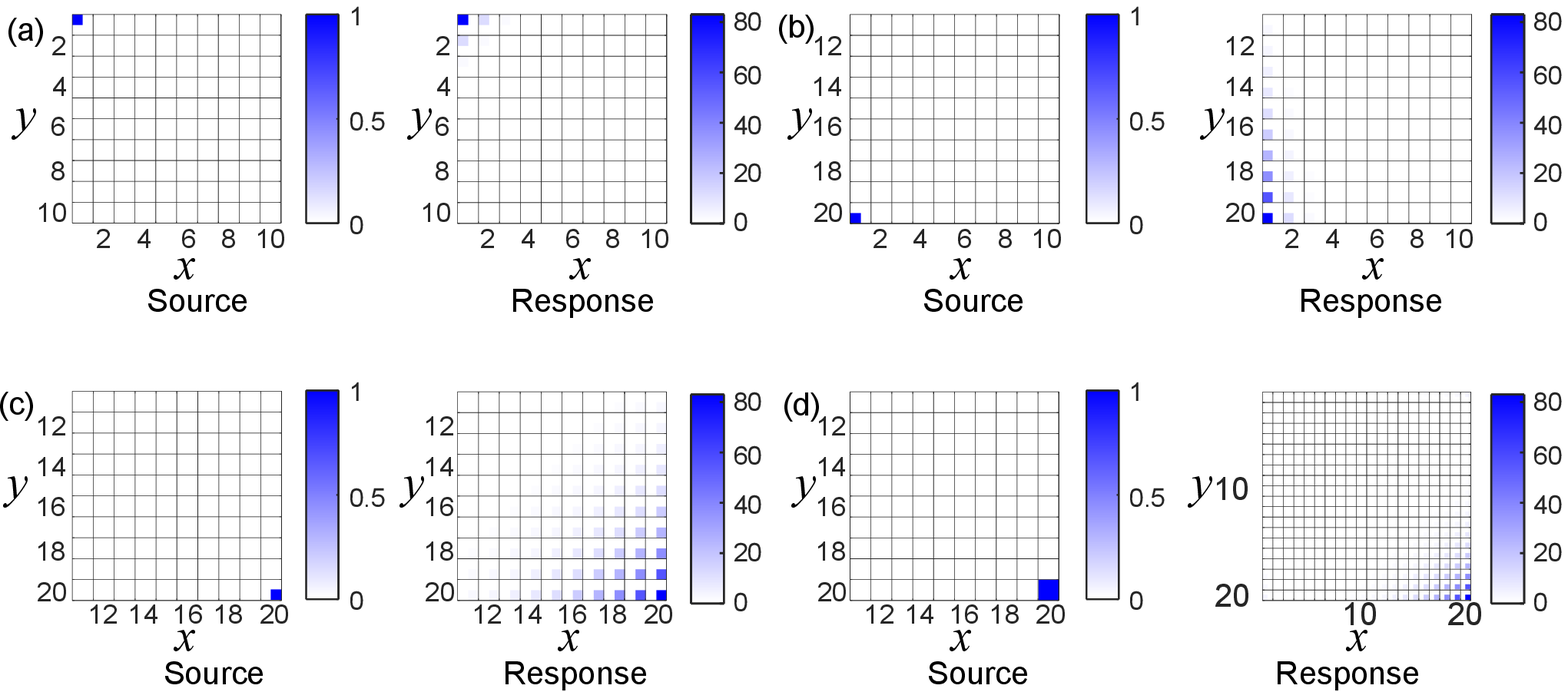}
\caption{The response of a non-Hermitian QI in near-Hermitian regime to external source at different corners. Color: magnitude of the complex field $\psi$. (a) Source is placed on $|x,y,j\rangle=|1,1,1\rangle$. (b) Source is placed on $|1,20,3\rangle$. (c) Source is placed on $|20,20,4\rangle$. (d) Sources of equal amplitude are placed on the four sublattices of the bottom-right unit cell, i.e., $|20,20,j\rangle,j=1,2,3,4$. Source frequency $E=0$, uniform on-site loss $\Gamma=0.01$. Other parameters: same as in Fig.~2 ($t=0.6, \gamma=0.4, \lambda=1.5, N=20$).}
\label{fig:driven}
\end{figure}

As mentioned in the main text, in the near-Hermitan regime the response of the system near zero energy is similar to that of a Hermitian QI. For completeness we show the numerical results here.  Fig.~\ref{fig:driven}(a-c) shows that, placing the source at a corner excites the corner state that is localized at that corner. This is because in this regime, the left and right eigenvectors of a corner state are localized at the same corner. In Fig.~\ref{fig:driven}(d) we put sources of equal amplitude on the four sublattices of the bottom-right unit cell, but the response is almost identical to Fig.~\ref{fig:driven}(c), showing that unlike in the intermediate regime, here it is practically impossible to excite an corner state that locates far away from the source.

\bibliographystyle{unsrt}
\bibliography{references}